# Epileptic Seizures: Quakes of the brain?

Ivan Osorio*[§], Mark G. Frei[§], Didier Sornette[¶], John Milton[#] & Ying-Cheng Lai**

\* Department of Neurology, The University of Kansas Medical Center, 3599 Rainbow Boulevard, Mailstop 2012, Kansas City, KS 66160, USA. email: iosorio@kumc.edu

[§] 5040 Bob Billings Parkway, Suite A, Lawrence, KS 66049, USA. email: frei@fhs.lawrence.ks.us

[¶] Department of Management, Technology and Economics, ETH Zurich, Kreuzplatz 5, CH-8032 Zurich, Switzerland. email: dsornette@ethz.ch

[#] W.M. Keck Science Center, The Claremont Colleges 925 N. Mills Avenue, Claremont, CA 91711, USA. Email: jmilton@jsd.claremont.edu

** Department of Electrical Engineering, Arizona State University, Tempe, AZ 85287-5706, USA. Email: ying-cheng.lai@asu.edu

**The concept of universality proposes that dynamical systems with the same power law behaviors are equivalent at large scales. We test this hypothesis on the Earth's crust and the epileptic brain, and discover that power laws also govern the distributions of seizure energies and recurrence times. This robust correspondence is extended over seven statistics, including the direct and inverse Omori laws. We also verify in an animal seizure model the earthquake-driven hypothesis that power law statistics co-exist with characteristic scales, as coupling between constitutive elements increases towards the synchronization regime. These observations point to the universality of the dynamics of coupled threshold oscillators for systems even as diverse as Earth and brain and suggest a general strategy for forecasting seizures, one of neurosciences' grails.**

Seemingly unpredictable, transiently-disabling seizures (SZ) are the hallmark of epilepsy, a neurological disorder that affects 60 million humans worldwide. SZ are typically associated with marked paroxysmal increases in the amplitude or rhythmicity of neuronal oscillations, which in a large number of subjects begin in a discrete region of the brain, but may eventually spread to engulf the entire brain. Although occurring in different media (the earth's crust, not the brain's cortex), earthquakes (EQ) also manifest as sudden, aperiodic, potentially damaging oscillations of morphology and spectra similar to seizures, albeit in different frequency ranges (Fig. 1). While these phenomena (SZ and EQ) are generated by vastly dissimilar systems, the aforementioned resemblances and others raise the possibility of similar dynamics, particularly since both SZ and EQ may be conceptualized as relaxation events, in which accumulated energy is discharged and/or transferred within a complex network of coupled threshold oscillators of relaxation (1). Using seven statistics, we investigate the extent to which these resemblances transcend into the phenomenological realm, by performing a quantitative comparison between catalogs of SZ recorded from humans with pharmaco-resistant epilepsies and of EQ within the Southern California Seismic Network (SCSN). We then expand this comparison and delve into shared mechanisms that may underlie the generation of these two phenomena by studying seizure dynamics in a maximally-driven model of generalized seizures (2). The evidence thus accumulated strongly supports the existence of a profound analogy between EQ and SZ organization and dynamics.

We performed quantitative analyses and comparisons of EQ observables in the SCSN catalog (81977 events with magnitude ≥ 2, 307019 total EQs, 1984-2000) with those of SZ (16032 detections) in continuous recordings directly from the brains of 60 human subjects with mesial temporal and frontal lobe pharmaco-resistant epilepsies (1996-2000; University of Kansas Medical Center). Recordings were performed on these subjects while they were maintained on reduced dosage of anti-seizure medications. SZ were defined in humans as the dimensionless ratio of brain electrical activity in a particular weighted frequency band reaching a threshold value, T, of at least 22 and remaining at or above T for at least 0.84 s (duration constraint, D) as previously described elsewhere (3,4). For experiments in rats, the SZ criteria were T=10 and D=0. The differences in T and D between humans and rats, reflect the attempt to maximize sensitivity and specificity of automated detection, according to accepted visual criteria for scoring SZ. The validated algorithm (3,4) for SZ detection also provided time of onset and termination, duration, site of origin, and intensity. These quantities were further processed to derive two key features: (1) "Energy" (E), defined as the product of SZ peak intensity and duration, which is considered by us to be the equivalent of EQ seismic moment ($S \sim 10^{1.5M}$, where M=magnitude); and (2) inter-SZ-interval (ISI), defined as the time elapsed between the onset of consecutive SZ, the equivalent of inter-quake-interval.

**"Gutenberg-Richter law" for SZ:** The probability density function (pdf) for SZ energy, E, was estimated using both histogram-based and adaptive Gaussian kernel-based methods. The pdf for the EQ seismic moment, S, was estimated in a similar manner. The pdf for E follows an approximate power law distribution, whose slope is statistically indistinguishable from the slope of the pdf for S (Gutenberg-Richter law). For both systems, the probability of an event having energy, E, for SZ (or seismic moment S for EQ) larger than x is proportional to $x^{-\beta}$, where $\beta \approx 2/3$ (Fig. 2).

**"Omori law" and "inverse Omori law" for SZ:** To investigate whether SZ behavior can be described by the Omori law for EQ, we time-locked the indicator function of being in SZ to each end time and averaged the indicator function at each time point. We then normalized the resulting curves by the total fraction of time each subject spent in SZ and averaged the resulting curves across all subjects. For the inverse Omori law, we followed these same steps, except that the indicator function was time-locked to each SZ onset time. The results (Fig. 3) indicate in humans an increased probability of other SZ occurring in the window beginning 30 minutes before a SZ and ending 30 minutes afterward. The same analysis method applied to the SCSN EQ catalog also showed both foreshock and aftershock probability increases, but on much longer time scales (over 25 days before and 30 days afterward). We purposefully avoided using any filtering or conditioning as done in standard seismic analysis (5-7), so as to mimic as closely as possible the procedure used for SZ. We are aware of the need for utilizing more sophisticated stacking methods for EQ and SZ in order to take into account space and time clustering, but we argue that the present approach has the advantage of recovering robust non-parametric evidence for EQ foreshocks and aftershocks, documented previously by more elaborate approaches (see reference [6] and references therein).

The observation of precursory events (equivalent to EQ foreshocks) in SZ time series is particularly striking, as it uncovers the potential for real-time forecasts, based on the detection of these precursors. The approximate symmetry between the "foreshock" and "aftershock" rates for human SZ, compared with the rather strong asymmetry in the rates of EQ foreshocks and aftershocks, may be interpreted, according to current models of triggered seismicity (8), as resulting from: 1) a thinner tail for the "Omori decay of SZ aftershocks" compared to EQ aftershocks rates; and 2) a weaker energy dependence of the triggering efficiency of future SZ by past SZ.

**Distribution of inter-event waiting times.** The pdf estimates for inter-event intervals were calculated and compared for EQ and SZ, using both histogram-based and adaptive Gaussian kernel-based estimation methods. The results (Fig. 4) illustrate that both densities approximately follow power law distributions, although with different slopes. For inter-quake intervals, we recover previous investigations (9) showing the survival distribution (probability that the inter-event time exceed $x$) is approximately proportional to $x^{-\beta}$, where $\beta \approx 0.1$, while for inter-seizure intervals, $\beta \approx 0.5$. We do not insist on the existence of genuine power laws for this statistic, as it has been shown that the pdf of inter-event times is probably more intricate than a pure power law (10). Our message is rather to stress the existence of a similar, heavy tail structure of this statistic in both EQ and SZ data.

**The expected time until the next event, conditioned on the time elapsed since the last one.** We test a prediction derived from the heavy tail structure of the waiting time distributions between successive events shown in Fig. 4. In the context of EQ, Davis et al. (11) showed the paradoxical result that, for such heavy tailed distributions, "the longer it has been since the last event, the longer the expected time till the next". Sornette and Knopoff (12) showed that, for a power law, the dependence of the average conditional additional waiting time until the next event, denoted $<\tau|t>$, is directly proportional to the time $t$ already elapsed since the last event. We test this prediction by computing $<\tau|t>$ empirically for each subject and compared it to its analogous statistic in the EQ SCSN catalog. Figure 5 shows indeed both for EQ and SZ that, for short times t since the last event, $\langle\tau|t\rangle$ is smaller than the (unconditional) average waiting time $\langle\tau\rangle$ between two events, and then increases until it becomes significantly larger than $\langle\tau\rangle$ as t increases. The saturation and subsequent decay after the peaks is due to finite size effects for both SZ and EQ.

**A test of the EQ-SZ analogy.** How is it possible that these strikingly different systems (brain and crust), operating at very different space and time scales, with markedly different underlying electro-mechanical-chemical processes, exhibit so many remarkable statistical similarities, as shown in Figs. 1-5? Several authors have speculated that EQ and SZ should indeed be similar, because they are both made of interacting threshold elements, and it has been recognized that such systems generically exhibit self-organized behavior with non-Gaussian statistics (13-15). However, the fact that EQ and SZ both possess a heavy-tailed distribution of event sizes should not be used *a priori* as sole support for such an analogy, because such one-point power law statistics can result from many distinct mechanisms (16). Here, we press this tenet further, informed by the fact that such systems tend to synchronize when the coupling strength between the elements is strong enough and/or the heterogeneity is weak enough (17-19). Thus, non-Gaussian statistics (sometimes attributed to self-organized criticality (20,14) co-exist with synchronized behavior in a general phase diagram in the heterogeneity-coupling strength plane (19). This led us to predict that manipulating, e.g., rats' brains so as to ensure strong neuronal coupling should generate a different class of statistics than reported above for humans, distinguished by characteristic time and size scales. We tested this prediction in 19 rats treated intravenously with the convulsant 3-mercapto-proprionic acid (3-MPA) at maximally tolerable (for viability) steady-state brain concentrations (2). Figure 6 shows the same statistics for rats as those reported in figure 2-5 for humans. The major difference between them, is the appearance of a clear "shoulder" in the rats' pdf of SZ energies (Fig. 6A) and of a peak around 13s in the pdf of inter-SZ-onset intervals (Fig. 6B). These two statistical features reveal the existence of a characteristic SZ size and of a quasi-periodic (synchronized) behavior, along with scale-free events. The quasi-periodicity is clearly seen in the SZ "foreshock/aftershock" diagram (Fig. 6C) in the shape of regularly spaced oscillations decorating the inverse and direct "Omori" laws. The average conditional waiting time (Fig. 6D) is also symptomatic of a quasi-periodic behavior superimposed with some large waiting time occurrences.

These findings imply that as with sand piles, block-spring Burridge-Knopoff and EQ-fault models, a power law regime (probably self-organized critical) "co-exist" with one of high synchronization, leading to events with a characteristic scale (*i.e.*, periodic). When coupling is strong, synchronization is generalized and seizures occur periodically; as the coupling strength decreases (lower 3-MPA concentrations), synchronization weakens and a power law regime (where severe SZ are infrequent) emerges. For these reasons, we label the mechanisms underlying the power law regime (SOC) found in humans (Figures 2-5) as "critical asynchronization" (21).

We have shown that the dynamical behavior of seizures, originating from discrete brain regions in subjects with pharmaco-resistant epilepsies on reduced doses of medications, may be described by scale-free power laws that are similar to those governing seismic activity. That is, intensity and duration are not defining properties of SZ and, contrary to the universally accepted practice in epileptology, these quantities should not be used as criteria to classify certain neuronal activity as either SZ or non-SZ. At a more abstract level, scale invariance in seizures may be conceptualized as the hallmark of certain complex systems, (the brain in this case) in which, at or near the critical point, its component elements (neurons) are correlated over all existing spatial , minicolum, column, macrocolum, etc.) and temporal scales (microseconds, seconds, tens of seconds, etc.) (16, 20). That the pdf of SZ energies E follows a power law, and more importantly that its exponent is $\beta \approx 2/3$ (as for EQ), has far-reaching, statistical-clinical implications: the mean and variance of E are mathematically infinite, which means in practice that the largest SZ in a given time series controls their values (3). As a consequence, variability is dominant and "typical" has no meaning. The energy pdf, and specifically its heavy tail, also suggests an explanation, at a mathematical-conceptual level, for the proclivity and capacity of the human brain to support status epilepticus, a potentially fatal condition characterized by prolonged/frequent SZ during which the brain does not return to its "normal" state, even when SZ activity abates.

The seven different statistics presented here and the rigorous validation procedure (22) of prediction, tested on maximally coupled rats' brains, provide strong support for a fundamental phenomenological analogy between SZ and EQ. At present, our results enable us to draw two important implications of the SZ-EQ analogy. First, extrapolating from SZ to EQ, we used the results obtained on strongly coupled rats' brains to posit that the controversial characteristic EQ hypothesis (23-25) is a regime for EQ organization that should be observed only when coupling between faults is strong and heterogeneity is weak (26) [see reference (27) for a balanced exposition of the "characteristic EQ" hypothesis and its implications for EQ prediction]. More generally, the theory underlying the correspondence between SZ and EQ suggests that wide spectra of different dynamic regimes are possible for systems such as the brain's cortex and the earth's crust. These regimes could correspond to critical asynchrony/SOC (20), clustering, quasi-periodicity, and/or synchronization, depending on the convulsant concentration, its rate of change, and other physico-chemical changes in the neuropil (engendered by SZ) that may be likened to changes in soil structure/composition/water content associated with EQ. Future tests of the SZ-EQ analogy should involve the question of seismic localization (faults) versus SZ "focus"/Epileptogenic Zone as conventionally defined, versus the concept of a distributed epileptic network and its propagation pathways.

Second, the evidence for both direct and inverse "Omori" laws for SZ and the long-memory associated with the heavy-tail distribution in inter-event times suggests the existence of genuine long-time interactions between SZ. In analogy with EQ, where this idea has taken shape under the concept of "triggering" (28) according to which most of the observed seismicity is probably due to past seismicity, this suggests a novel research direction for the prediction of SZ based on the notion that SZ beget SZ.  In seismology, it has been recognized that the many small, undetected EQ provide a major if not dominant contribution to the triggering future of EQ of any size (7). Prolonged recordings of brain cortical electrical activity (ECoG), the equivalent of seismographs, from epileptic humans and animals contain frequent, low intensity, short bursts of abnormal activity unperceived by the patient and observers and interspersed with infrequent, but longer, more widespread, and more intense bursts (convulsions) (4). The SZ-EQ analogy, including the evidence presented here for an inherent capacity of SZ to trigger future SZ, suggest that a workable prediction scheme should use the triggering by, not only past perceived (clinical) SZ, but also the myriad of unperceived (subclinical) abnormal neuronal bursts.

The totality of our findings justifies a novel approach to forecasting SZ that encompasses not only their intrinsic triggering capacity, but expands the set of monitored observables from the local (epileptogenic zone/focus) to the global (epileptic network) and from clinical seizures to all types of epileptiform activity (subclinical seizures and other related paroxysmal oscillations), while taking into account the prevailing epileptic state (e.g., critical asynchrony vs. quasi-periodicity) and the system's history at the time of the forecast. This strategy may bring us closer to one of the "grails" of neuroscience: the prediction and prevention of SZ in humans.

29. The steady-state 3-MPA rat model was developed with C. Lunte and E. Crick from the Ralph M. Adams Institute of Bioanalytical Chemistry, Univ. of Kansas and the rat seizure data was collected in C. Lunte's lab. N. Bhavaraju performed preliminary data analysis. This work was supported in part by NIH/NINDS grant nos. 5R21NS056022 and 1R01NS046602. In Memory of Boris Birmaher Ghitis, M.D.

**Figure Legends**

**Fig. 1**: (color online)  (**A**) Human electrocorticogram, recorded directly from the brain, containing a SZ and a "foreshock". Data were sampled at 240 Hz; 150 s of data are shown. The second derivative of the signal ("acceleration") is displayed. (**B**) Vertical acceleration recorded during the October 17, 1989 Loma Prieta earthquake in the Santa Cruz Mountains. Data were sampled at 200 Hz; 30 s of data are shown. (**C**) Power spectral density estimates for 20 SZ (2 each from 10 subjects) and for the 3 triaxial acceleration recordings from the Loma Prieta EQ.

**Fig. 2**: (color online) Probability density functions of seismic moments from the SCSN EQ catalog from 1984 to 2000 (blue) and of SZ energies (red). Both statistics are compatible with the same power law with exponent ≈ 2/3. The scaling range for SZ is smaller than for EQ, probably due to the marked size difference between the human brain cortex and the earth's crust.

**Fig. 3**: (color online) Superimposed epoch analysis of EQ and SZ to test in SZ for the existence of aftershocks ("Omori law") and foreshocks ("inverse Omori law"). In SZ (red top-right scales), the presence of aftershocks was investigated using the indicator function of being in SZ, time-locked to SZ end-time and then averaging at each time point. The curves were normalized by the fraction of time each subject spent in SZ and then averaged across all subjects.  For foreshocks, the same procedure was followed, except that the indicator function was time-locked to SZ onset-time. For EQ (blue bottom-left scales), all sequences of EQ with magnitude ≥ 2 in the SCNS catalog preceding and following EQ of magnitude ≥ 5 are stacked.

**Fig. 4**: (color online) Probability density function estimates of inter-event intervals for SZ (red upper-right scales) and EQ (blue lower left scales). For EQ, we used return times for quakes with M≥3 and epicenters within the same cell of a one degree grid in the SCSN catalog. For SZ, we used inter-detection-interval for all events, regardless of onset location.

**Fig. 5**: (color online) Average conditional waiting time <|t> until the next event conditional on the fact that a time t has already elapsed since the last event. The red top-right scales correspond to the SZ statistics, while the blue bottom-left scales correspond to the EQ statistics. One can observe for both SZ and EQ the paradoxical behavior that <|t> increases with t up to a maximum (due to finite size effects). The dashed horizontal line shows the value of the unconditional average waiting time between two events. The increases of <|t> with t confirm the heavy-tailed nature of the distribution of inter-event times.

**Fig. 6**: (color online) Same statistics as those reported for humans in figures 2-5 but in this case for 19 rats treated intravenously (2) with the convulsant 3-mercapto-proprionic acid (3-MPA). **(A)** The pdf of SZ energies exhibits both a power law and a characteristic size (see arrow). **(B)** The pdf of inter-SZ-onset interval shows a power law and a peak indicative of quasi-periodicity (see arrow at approximately 13 s). **(C)** The "foreshock-aftershock "diagram exhibits the inverse and direct Omori laws, decorated by oscillations with a period of 13 s. **(D)** The expected waiting time, conditioned on time since last seizure, is symptomatic of the presence of both quasi-periodic and SOC regimes within the same system.

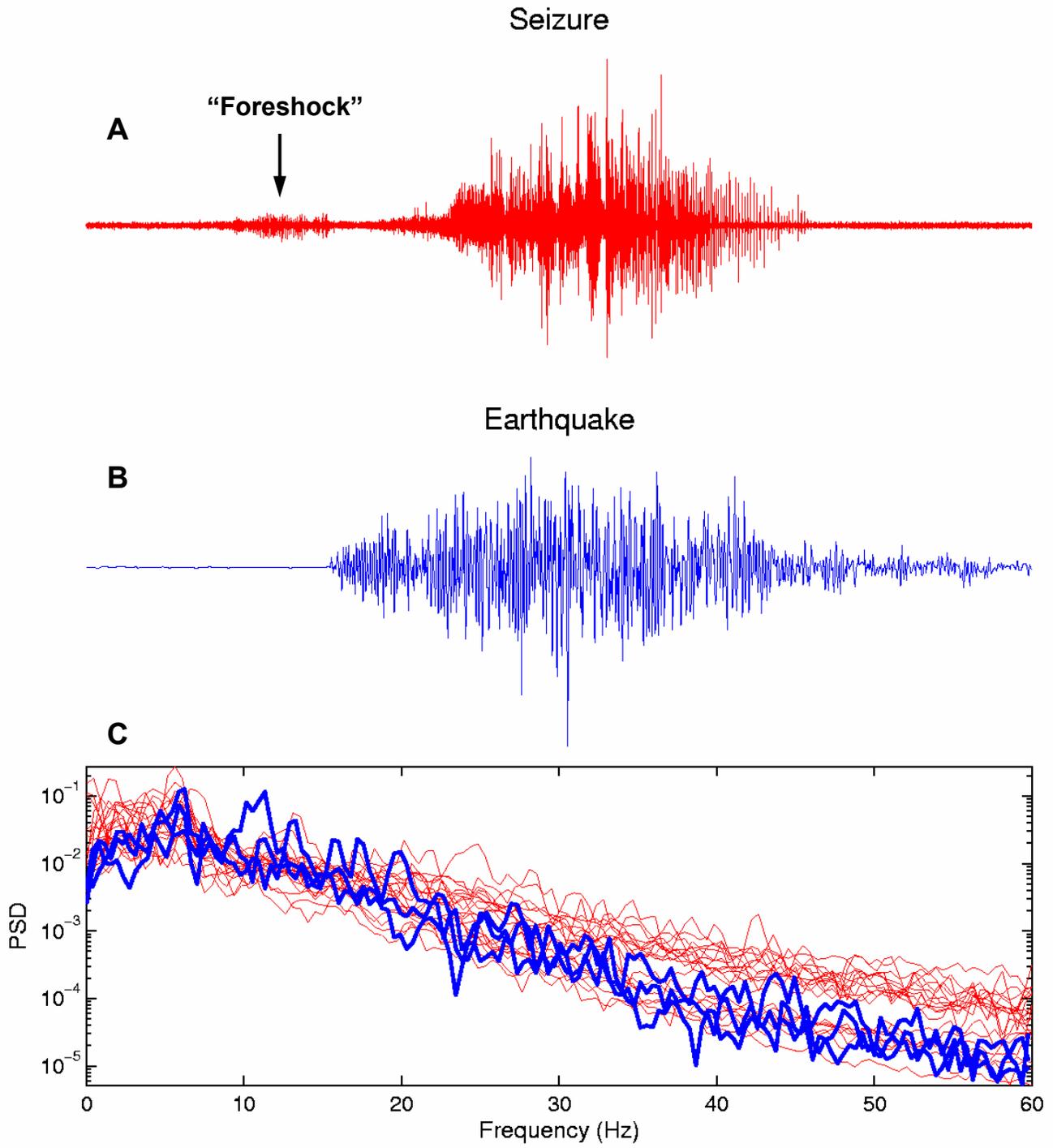

**Figure 1**

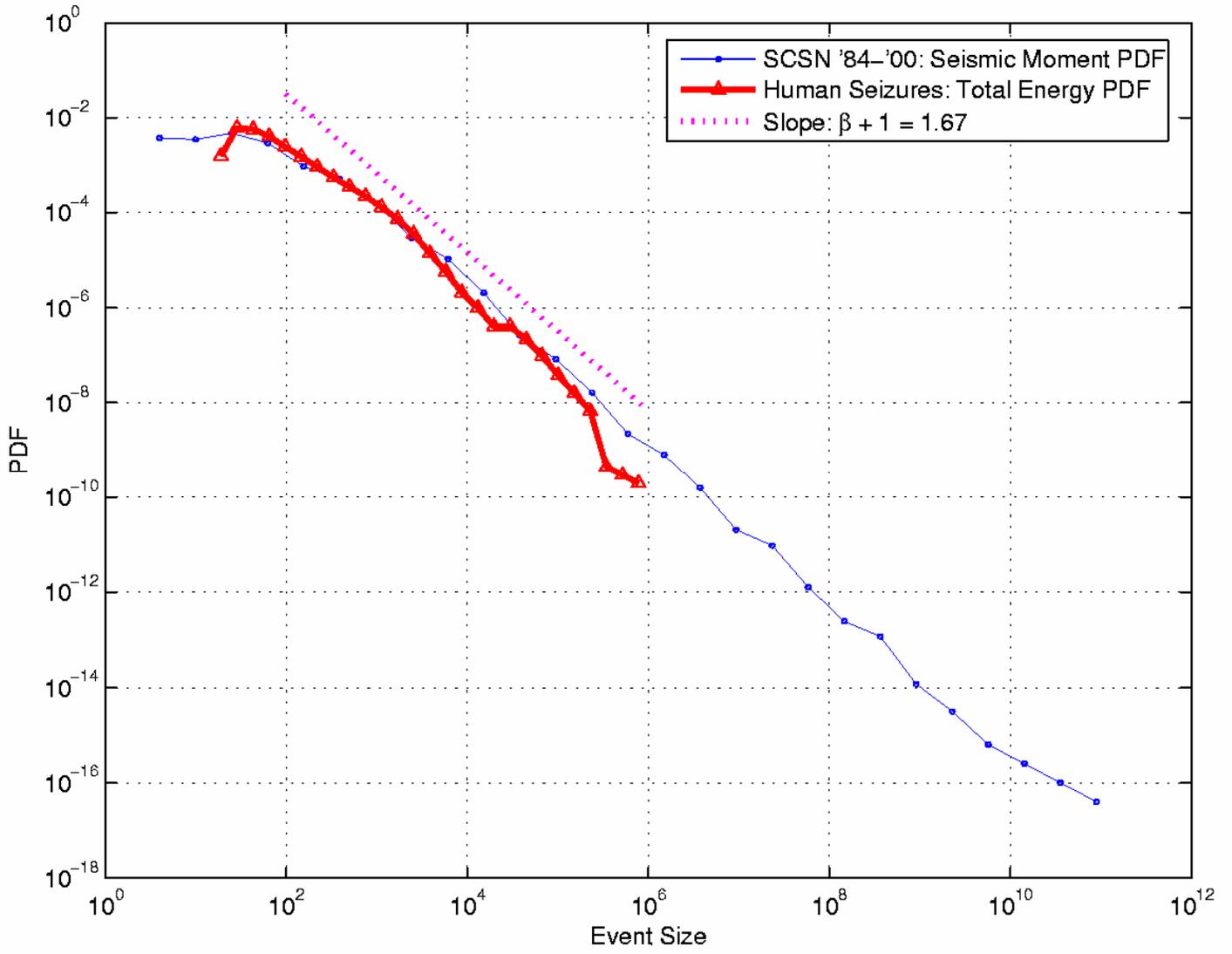

**Figure 2**

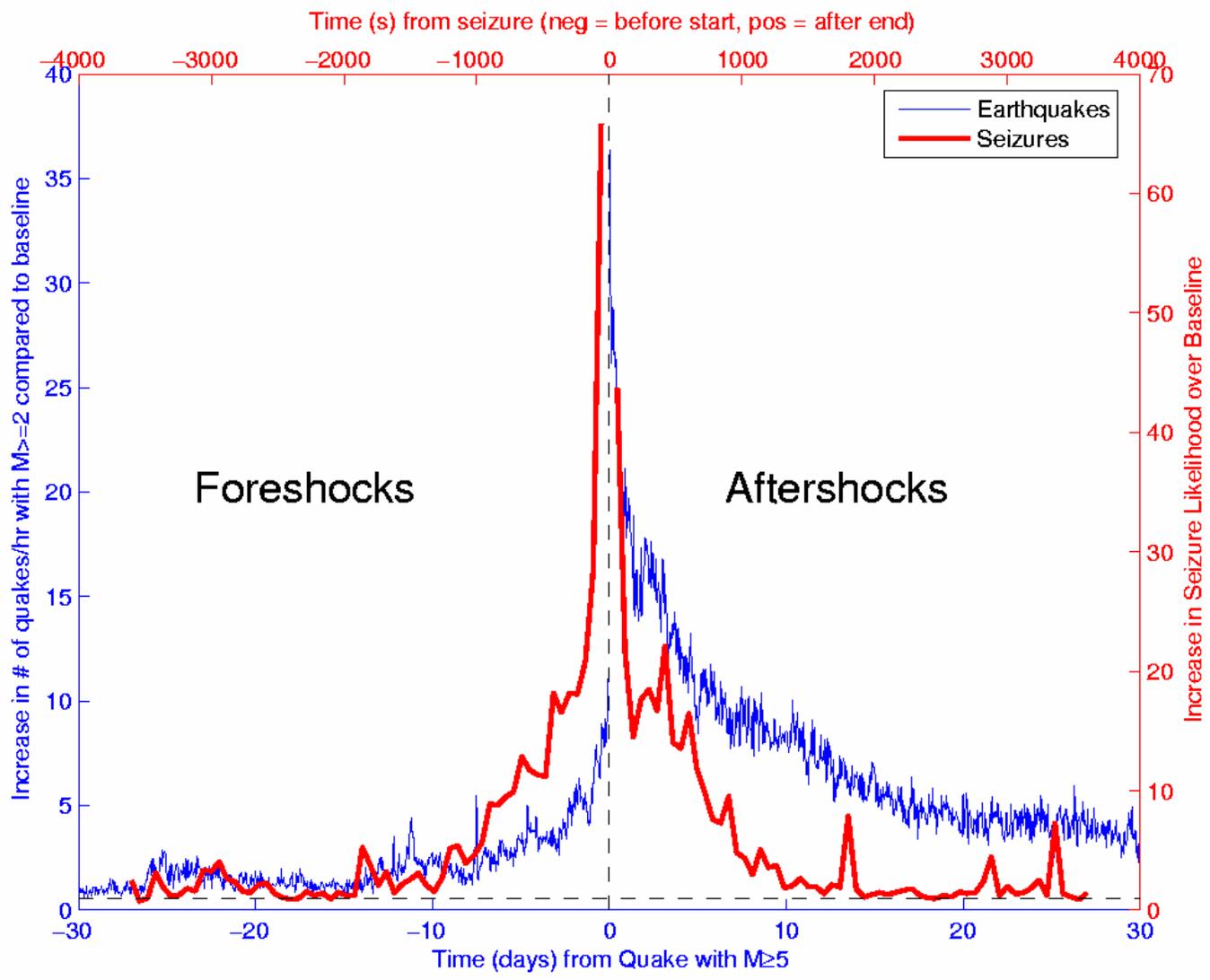

**Figure 3**

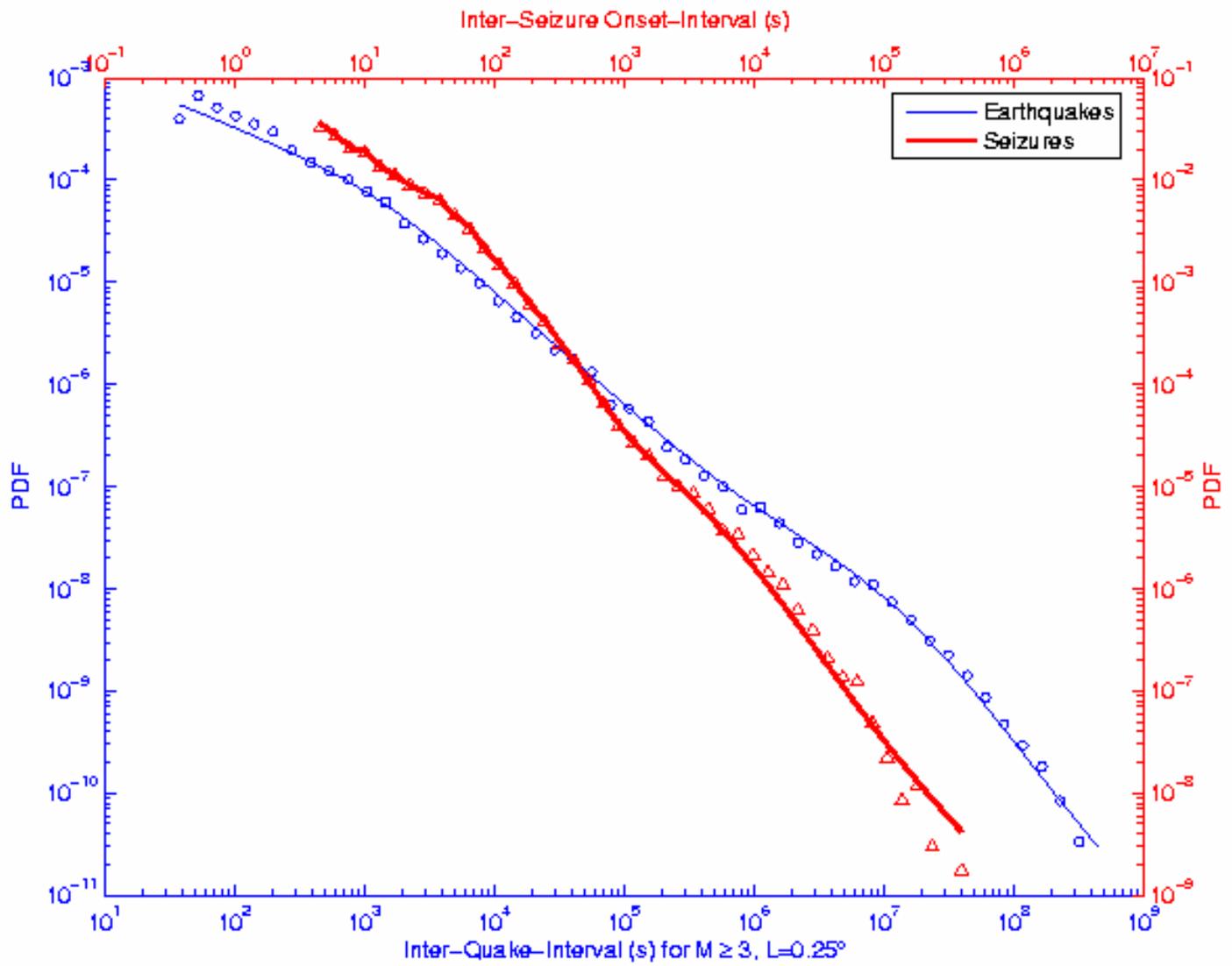

**Figure 4**

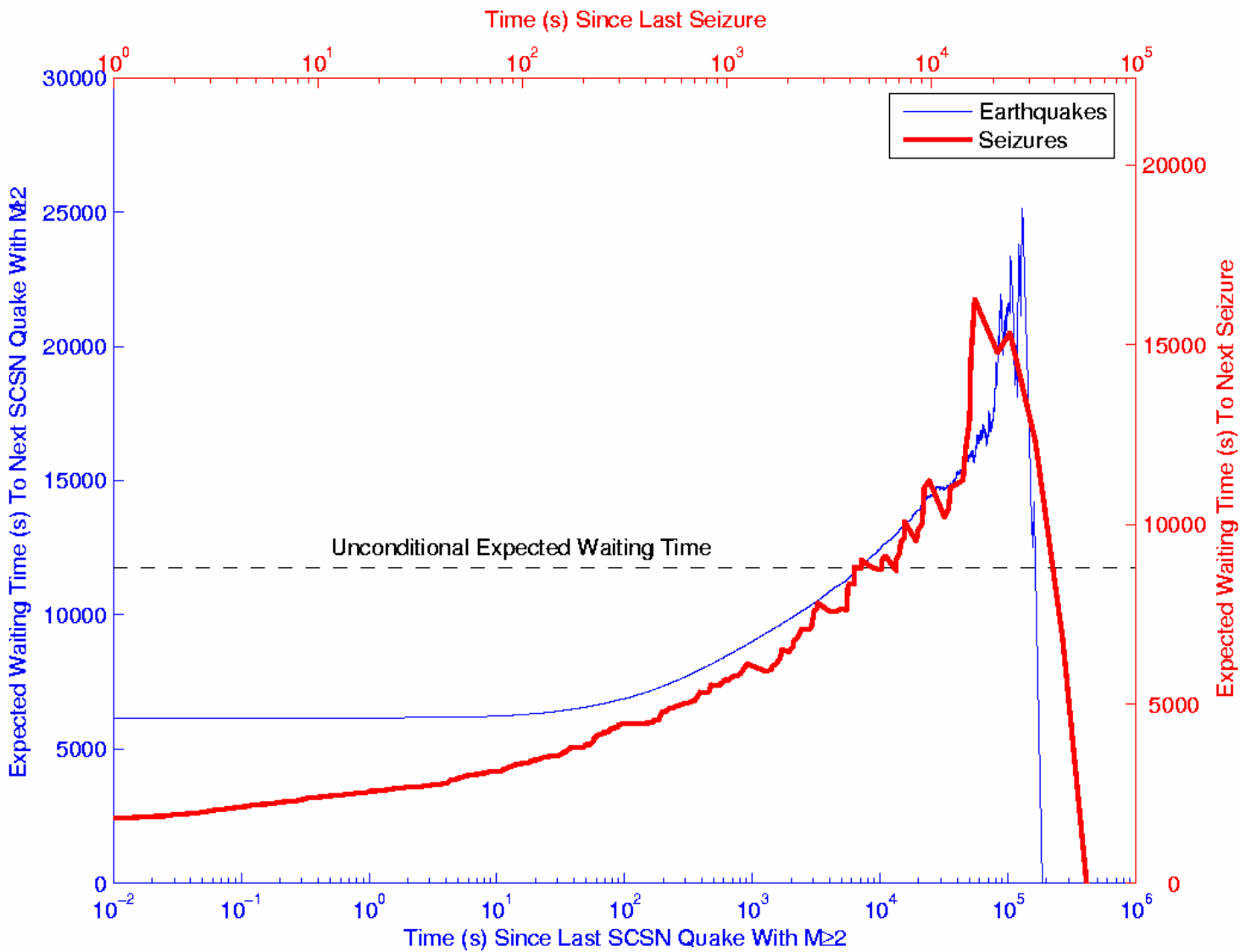

**Figure 5**

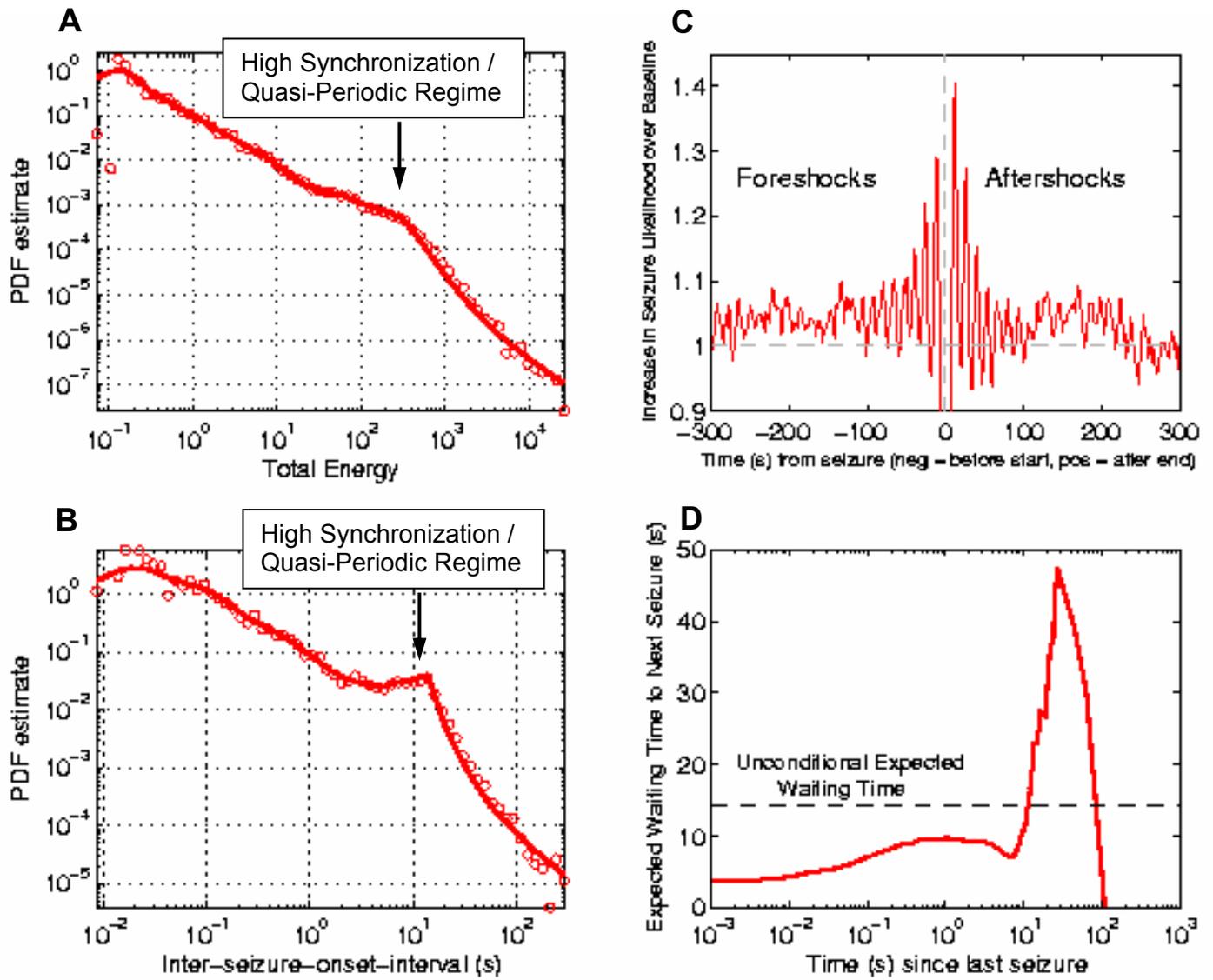

**Figure 6**